**Finding Efficient Collective Variables: The Case of Crystallization**

Yue-Yu Zhang[1,2], Haiyang Niu[1,2], GiovanniMaria Piccini[1,2], Dan Mendels[1,2], and Michele Parrinello[1,2]*

1   Department of Chemistry and Applied Biosciences, ETH Zurich, c/o USI Campus, Via Giuseppe Buffi 13, CH-6900, Lugano, Ticino, Switzerland

2   Institute of Computational Science, Università della Svizzera italiana (USI), Via Giuseppe Buffi 13, CH-6900, Lugano, Ticino, Switzerland

*E-mail: parrinello@phys.chem.ethz.ch

**Abstract**

Several enhanced sampling methods such as umbrella sampling or metadynamics rely on the identification of an appropriate set of collective variables. Recently two methods have been proposed to alleviate the task of determining efficient collective variables. One is based on linear discriminant analysis, the other on a variational approach to conformational dynamics, and uses time-lagged independent component analysis. In this paper, we compare the performance of these two approaches in the study of the homogeneous crystallization of two simple metals. We focus on Na and Al and search for the most efficient collective variables that can be expressed as a linear combination of X-ray diffraction peak intensities. We find that the performances of the two methods are very similar. However, the method based on linear discriminant analysis, in its harmonic version, is to be preferred because it is simpler and much less computationally demanding.

## 1. Introduction

Many chemical and physical phenomena are characterized by the occurrence of long lived metastable states. Under these circumstances, accurate sampling becomes very computationally expensive or even prohibitive. In order to overcome this problem many methods have been suggested [1]. A large fraction of these methods depends on the definition of collective variables (CVs). Typical examples are umbrella sampling [2-5], metadynamics [6-9] and variationally enhanced sampling [10]. The efficiency of these simulations depends very much on the quality of the CV, and hence the finding and improving of CVs is the object of intense investigations [11].

In our group two methods have recently been developed, harmonic linear discriminant analysis (HLDA) [12] and variational approach to conformational dynamics (VAC) [13]. In HLDA one constructs low dimensional CVs from the local fluctuations in the different metastable states. In contrast in the VAC approach one starts with a biased simulation with non-optimal CVs and attempts to improve this initial guess using a variational principle approach that is based on the time-lagged independent component analysis (TICA).

In this work we compare the performance of HLDA-generated CVs with those obtained using VAC. We focus on a typical application area of enhanced sampling methods, namely homogeneous crystallization. In particular we shall present results on two systems, Na and Al, already studied elsewhere [4,5,9,14].

It has recently been shown [15] that the X-ray diffraction (XRD) intensities can be useful CVs. However, sometimes a single peak is not sufficient. Thus we shall search with HLDA and VAC for the best linear combinations of the diffraction peak intensities.

## 2. Methods

## 2.1 Well tempered metadynamics

Here, we use as an enhanced sampling method well-tempered metadynamics (WTMetaD) [16, 17]. We recall that WTMetaD is a procedure in which the dynamical evolution of the system is altered by the addition of an external bias potential $V(s)$ periodically updated as:

$$V_n(s) = V_{n-1}(s) + G(s, s_n) \exp\left[-\frac{1}{\gamma - 1}\beta V_{n-1}(s_n)\right], \qquad (1)$$

where $\gamma$ is the bias factor and $\beta = 1/k_B T$ is the inverse temperature. The modification of the potential in eq. 1 results from the multiplication of a Gaussian kernel $G(s, s_n)$ centered at the current CV value $s_n$ and scaled by $\exp\left[-\frac{1}{\gamma-1}\beta V_{n-1}(s_n)\right]$. This scaling factor decreases as $1/n$, thus the change of the external bias potential becomes smaller as the metadynamics simulation progresses. The bias potential $V(s(R))$ asymptotically takes the form

$$V(s) = -\left(1 - \frac{1}{\gamma}\right)F(s), \qquad (2)$$

where $F(s)$ is the free energy associated to the CV $s$. The validity of eq. 2 has recently been rigorously proven [18].

## 2.2 Harmonic linear discriminant analysis

Very recently a simple way of obtaining efficient CVs called harmonic linear discriminant analysis (HLDA) has been proposed [12]. In its simplest version one assumes that there are two separated states that can be identified by $N_d$ descriptors $d_i(R)$ that are a function of the atomic coordinates $R$ and are arranged to form a vector $d(R)$. We assume that the average values in the two basins, $\mu_A$ and $\mu_B$, are different and there is no overlap in the $N_d$ space between the two states. The fluctuation of $d(R)$ in the two states are given by the covariance matrices $\Sigma_A$ and $\Sigma_B$.

The idea of HLDA is to find the direction along which the projected data of the two states is best separated. This is obtained by maximizing the ratio of the between-class variation $S_B=(\mu_A-\mu_B)(\mu_A-\mu_B)^T$ and the within-class variance calculated by a harmonic average $S_W = \Sigma_A \Sigma_B/(\Sigma_A + \Sigma_B)$, which leads to the object function:

$$J(W) = \frac{W^T S_B W}{W^T S_W W}, \tag{3}$$

where $J(W)$ is maximized by

$$W = \left(\frac{1}{\Sigma_A} + \frac{1}{\Sigma_B}\right)(\mu_A - \mu_B). \tag{4}$$

Thus the HLDA based CV takes the form

$$s^H(R) = W^T d(R) = (\mu_A - \mu_B)^T \left(\frac{1}{\Sigma_A} + \frac{1}{\Sigma_B}\right) d(R). \tag{5}$$

This procedure has been inspired from linear discriminant analysis (LDA). However, it differs from the standard version of LDA in that the within-class variance is estimated from the harmonic average of the covariance matrices, and not by the arithmetic average as done in LDA. The reason for this choice has been illustrated in ref. 12.

An extension of HLDA to the case in which several metastable states need to be considered is possible and following the nomenclature of the artificial intelligence community we refer to this as a multi class situation and thereby we shall name the method as multi-class HLDA (MC-HLDA) [19]. We assume that there are $c$ classes, the between-class variance matrix in this case is expressed as $S_B = \sum_{i=1}^{c}(\mu_i - \mu)(\mu_i - \mu)^T$, where $\mu = \frac{1}{c}\sum_{i=1}^{c}\mu_i$, and the within-class variance matrix is expressed as $S_W = 1/((1/\Sigma_1)+(1/\Sigma_2)+\ldots+(1/\Sigma_c))$. The projection matrix $\mathbf{K} = [W_1|W_2|\ldots|W_c]$ that maximizes the ratio $J(\mathbf{K}) = \frac{\mathbf{K}^T S_B \mathbf{K}}{\mathbf{K}^T S_W \mathbf{K}}$ is the one whose columns are the eigenvectors corresponding to the generalized eigenvalue problem

$$(S_B - \lambda_i S_W)W_i = 0. \tag{6}$$

The optimized CVs are thus given by:

$$s_i^V(R) = W_i^T d(R). \tag{7}$$

Notice that, because $S_B$ is of rank ($c$-1) or less, there will be at most $c$-1 eigenvectors with non-zero eigenvalues $\lambda_i$.

### 2.3 Variational approach to conformational dynamics

Another way one can use to optimize CVs is based on a variational approach to conformational dynamics (VAC) [13]. It has been shown that time-lagged independent component analysis (TICA) can provide an optimal solution of VAC. TICA is a well-known method for blind source separation problems in signal processing, and recently have been applied in finding slow modes in molecular dynamics simulations [20-26].

The first version of this method required the knowledge of the trajectory [21, 22]. However recently this method has been extended and it is now possible to perform a VAC analysis starting from biased simulations with non-optimal CVs [13]. These CVs are then optimized. In the application of this principle one expand the slow modes into a basis set of $N_d$ descriptors $\boldsymbol{d_i}(\boldsymbol{R_t})$ ($i$ = 1, 2, … , $N_d$). The descriptors are normalized to one and the average value is subtracted to get a quantity of zero mean. The best linear combination in these basis sets is obtained by maximizing the ratio $f(\boldsymbol{W_1})$ with respect to a projection vector $\boldsymbol{W_1}$:

$$f(\boldsymbol{W_1}) = \frac{\boldsymbol{W_1^T}|\overline{\boldsymbol{C}}(\tau)|\boldsymbol{W_1}}{\boldsymbol{W_1^T}|\overline{\boldsymbol{C}}(0)|\boldsymbol{W_1}}, \tag{8}$$

where $\boldsymbol{C_{ij}}(\tau) = E[\boldsymbol{d_i}(\tilde{t})\boldsymbol{d_j}(\tilde{t}+\tau)]$ is the time lagged correlation between $\boldsymbol{d_i}$ and $\boldsymbol{d_j}$, $\overline{\boldsymbol{C}}(\tau) = [\boldsymbol{C}(\tau) + \boldsymbol{C^T}(\tau)]/2$ is the symmetrized correlation matrix, and $\overline{\boldsymbol{C}}(0)$ is the correlation at lag time 0. In ref. 13 and 24 it has been shown that eq. 8 can be applied also to biased simulations provided that the time scale is modified as follow:

$$d\tilde{t} = e^{\beta(V(s,t)-c(t))}dt, \tag{9}$$

where $V(\boldsymbol{s},t)$ is the instantaneous value of bias in WTMetaD, and

$$c(t) = -\frac{1}{\beta}\log\frac{\int d\boldsymbol{s}\, e^{-\beta(F(s)+V(s,t))}}{\int d\boldsymbol{s}\, e^{-\beta F(s)}} \tag{10}$$

is an energy offset and $F(\boldsymbol{s})$ is the free energy of the system. The function $c(t)$ asymptotically tends to the reversible work done on the system by the external bias.

Maximizing the ratio in eq. 8 leads to the generalize eigenvalue problem:

$$\overline{C}(\tau)W_1 = \lambda(\tau)_1 \overline{C}(0)W_1. \tag{11}$$

Subsequent projections, $W_2, \ldots W_{N_d}$, can be similarly found by solving the same eigenvalue problem $\overline{C}(\tau) \cdot W = \lambda(\tau)\overline{C}(0) \cdot W$. The eigenvalues of eq. 11 can be ordered according to the $\lambda(\tau)$ values. The slowest decreasing modes are then chosen as CVs:

$$s_i^V(\boldsymbol{R}) = \boldsymbol{W}_i^T \boldsymbol{d}(\boldsymbol{R}). \tag{12}$$

**2.4 X-ray diffraction intensities**

In a recent paper [15] it has been suggested that the intensities of the X-ray diffraction (XRD) peaks could act as good CVs since they are physically meaningful and can distinguish between different states.

The XRD intensities are computed by the Debye scattering function:

$$I(Q) = \frac{1}{N}\sum_{i=1}^{N}\sum_{j=1}^{N} f_i(Q)f_j(Q)\frac{\sin(Q \cdot R_{ij})}{Q \cdot R_{ij}} W(R_{ij}), \tag{13}$$

where $N$ is the total number of atoms, $Q$ is the modulus of the scattering vector, $f_i(Q)$ and $f_j(Q)$ are the atomic scattering form factors, $R_{ij}$ is the distance between atoms $i$ and $j$, and $W(R_{ij})=\sin(\pi R_{ij}/R_c)/(\pi R_{ij}/R_c)$ is a window function that handles the problem of finite simulation box and $R_c$ is the upper limit of $R_{ij}$.

**3. Computational details**

All simulations were performed in isothermal-isobaric (NPT) ensemble. We simulated Na and Al using embedded atom models [27, 28]. Biased metadynamics simulations were performed using LAMMPS [29] patched with PLUMED 2 [30]. The integration of the equations of motion was carried out with a time step of 2 fs. We employed the stochastic velocity rescaling thermostat [31] with a relaxation time of 0.1 ps. The target pressure of the barostat was set to the atmospheric value and a relaxation time of 10 ps was used. In order to determine the free energy surface $F(\boldsymbol{s})$ as a function of the CVs, we used WTMetaD [16]. The temporal length of the WTMetad runs range from 600 to 800 ns. Details of the calculations can be found in the Supporting Information (SI) [32] (Table S1-S4).

We simulated the two systems at the temperature close to their melting points (375 K for Na and 850 K for Al). The total numbers of atoms were 250 in Na and 256 in Al.

## 4. Results and Discussion
### 4.1 Choosing the descriptors

Both HLDA and VAC require defining an adequate set of descriptors. As mentioned in the introduction we shall use some of the XRD peaks intensities. The cutoff $R_c$ for Na and Al are and 11 Å and 9 Å, respectively. Due to the small system sizes the peaks resulting from eq. 13 are broad as shown in Figure 1, but they can still distinguish well between solid and liquid.

In order to perform our analysis, we found that the intensities of the first four Bragg peaks were sufficient for our purpose. The four descriptors are labeled by the peaks Miller indices {hkl}. They are $d_1 = I^{\{011\}}$, $d_2 = I^{\{002\}}$, $d_3 = I^{\{112\}}$ and $d_4 = I^{\{022\}}$ in Na, and $d_1 = I^{\{111\}}$, $d_2 = I^{\{002\}}$, $d_3 = I^{\{022\}}$ and $d_4 = I^{\{113\}}$ in Al.

In order to make a meaningful comparison of the HLDA and VAC coefficients from the original runs in which only $I^{\{011\}}$ for Na and $I^{\{111\}}$ for Al is biased, we rescaled the value $I^{\{hkl\}}$ to $\tilde{I}^{\{hkl\}}$, which covers the range [0, 1] where 0 is the minimum value sampled and 1 is the maximum.

### 4.2 The Na case

From the existing literature it is known that the $F(s)$ of Na close to 375 K is dominated by 2 minima corresponding to the disordered liquid state and the ordered body centered cubic (BCC) states (Figure 2a).

We first applied HLDA to find the direction that best discriminate the two states. In each of them we estimate the average value and the covariance matrices in short unbiased runs. We then applied eq. 5 to obtain the CV, and the result is shown in Table 1. The descriptor $I^{\{011\}}$, which corresponds to the highest peak in the XRD pattern, dominates in the HLDA coefficients.

The VAC based calculation was started with a biased run in which only $I^{\{011\}}$ is used as CV. Following the procedure described in Ref. 13, we extract the slowest decaying modes in the basis of the selected descriptors. The eigenvalue of the VAC equation are displayed in Figure 2b as a function of lag-time. Obviously one eigenvalue dominates the long time decay. The coefficients for the descriptors are shown in Table 1. They are very similar to those of HLDA and also dominated by $I^{\{011\}}$.

The probability distributions related to these two new CVs were computed and shown in Figure 2c. Both of them can discriminate the two states, while the transition region is better separated by the direction of the HLDA CV $s^H$ than that of the VAC CV $s^V$.

We now compare the performance of $I^{\{011\}}$, $s^H$ and $s^V$ as a metadynamics CVs. Thus two additional runs based on $s^H$ and $s^V$ were performed. In Fig. 3, it is obvious that both $s^H$ and $s^V$ improves substantially the performance of the original simulation. It is hard to tell which one between $s^H$ and $s^V$ is more efficient. Judging only from the number of transitions per unit time $s^H$ appears to be slightly more efficient.

**4.2 The Al case**

The other case that we studied was the crystallization of Al. We first explored the free energy surface biasing $I^{\{111\}}$ at 850 K, close to the melting temperature. Following the reweighting procedure in ref. 33, we obtain the FES projected onto $d_1$ and $d_4$, as shown in Figure 4a. We chose this particular projection among all the possible pairs of descriptors because in this plane the states are more clearly distinguishable.

Beside the liquid and face centered cubic (FCC) phases a third minimum appears that correspond to a close packed (CP) structure in which disorderly arranged stacking faults appears. Such a phase of Al has indeed been found in the experiments [34-36].

This calculation suggests that these three different states should be considered. Thus, we apply the extension of HLDA: MC-HLDA in multi-class version [19]. To prepare for the MC-HLDA simulation we performed three independent 2-ns-long runs start from the three states, and on this time scale no transition between these states was observed. Thus we could estimate the mean value of the descriptors in all three states and their covariance matrix. We then apply eq. 6 and solve the generalized eigenvalue problem. We find two non-zero eigenvalues and the corresponding eigenvectors span a 2-dimensional space in which the three states are well discriminated. The values of these eigenvectors are reported in Table 2.

We then constructed a VAC calculation starting from the biased trajectory described earlier. It is obvious to see that in agreement with the MC-HLDA results, two eigenvalues dominate the long time decay (Fig. 4b), leading also to a 2-dimensional CV space. The coefficients of the corresponding eigenvectors are listed in Table 2.

Both the HLDA CV $s^H$ and the VAC CV $s^V$ spaces (Fig. 4c and d) can discriminate well between the three basins. While the coefficients of the descriptors appear to be different, the two FES are very similar with the three minima well separated, apart from a slight relative rotation.

To compare the relative performance of the $s^{\{111\}}$, $s^H$ and $s^V$ we performed two extra simulations using $s^H$ and $s^V$. We find again the performance of $s^H$ and $s^V$ to be superior to $s^{\{111\}}$ but otherwise very close.

## 5. Conclusion

In this paper, we tested the performance of HLDA and VAC in the case of homogenous nucleation. We applied them to look for the most efficient collective variables that can be expressed as a linear combination of X-ray diffraction peak intensities. Both methods gave a good separation among the different states and substantially enhanced sampling. However VAC involves a more complex and lengthy analysis. Thus our conclusion is that HLDA, for the problem here exhibited, is to be preferred.

From the theoretical point of view VAC is more soundly based as a method for calculating CVs. The fact that HLDA performs similarly well appears to be a validation of this more empirical approach.

**Acknowledgements**

The authors thank Faidon Brotzakis, Yi Issac Yang, Valerio Rizzi and Pablo M. Piaggi for useful discussions. Some of the calculations reported here were performed on the EULER cluster at ETH Zurich and the others were on the Mönch cluster at the Swiss National Supercomputing Center (CSCS). This research was supported by the VARMET European Union Grant ERC-2014-ADG-670227.

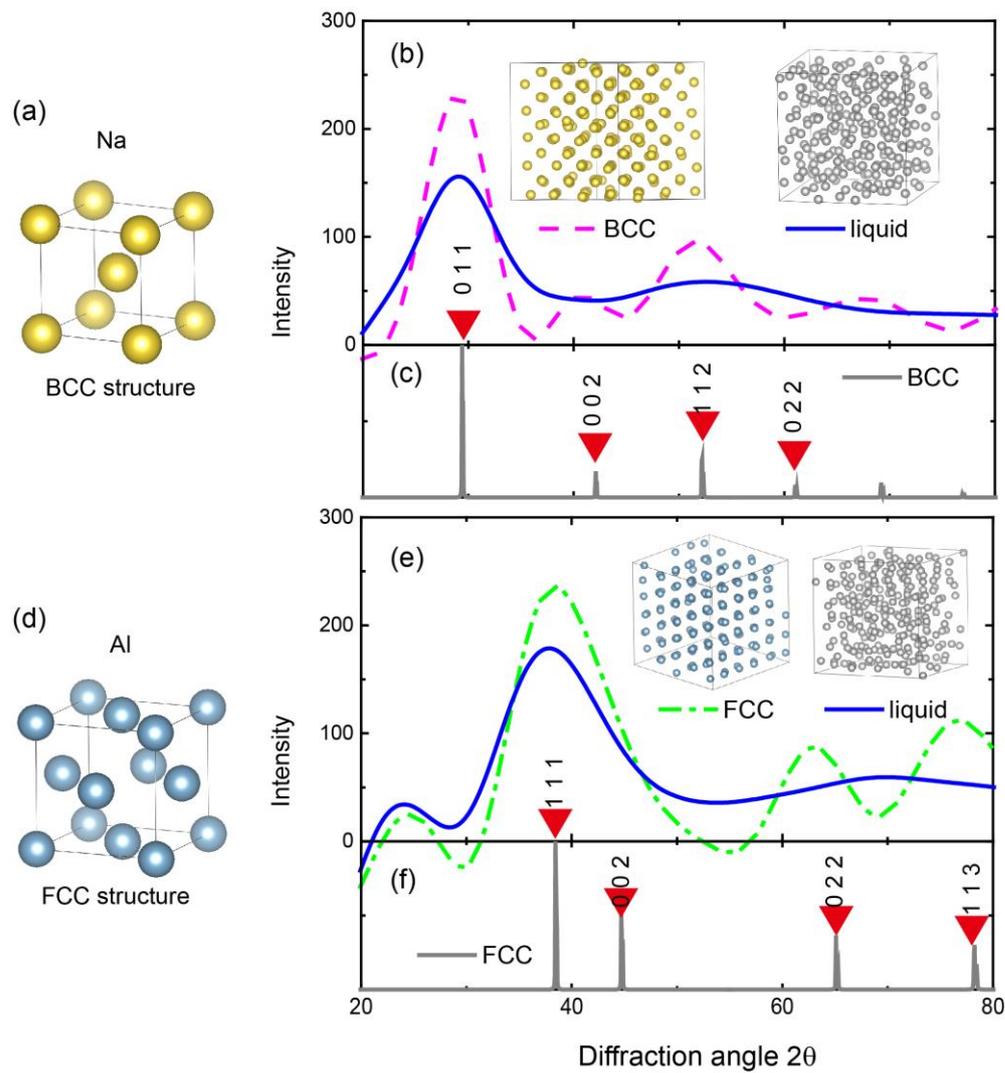

**Figure 1.** (a) Na unit cell in the BCC phase. (b) Simulated XRD patterns of BCC and liquid Na at 375 K with 250 atoms. The cutoff radius $R_c$ is set to 11 Å. Inset, two snapshots of the BCC (left) and liquid (right) phases. (c) Simulated XRD pattern for ideal Na BCC structure with lattice constant 4.29 Å. The first four peaks are highlighted in red triangles. (d) Al unit cell in the FCC phase. (e) Simulated XRD patterns for FCC and liquid Al at 850 K with 256 atoms. The cutoff radius $R_c$ is set to 9 Å. Inset, two snapshots of the FCC (left) and liquid (right) phases. (f) Simulated XRD pattern of ideal Al FCC structure with lattice constant 4.05 Å.

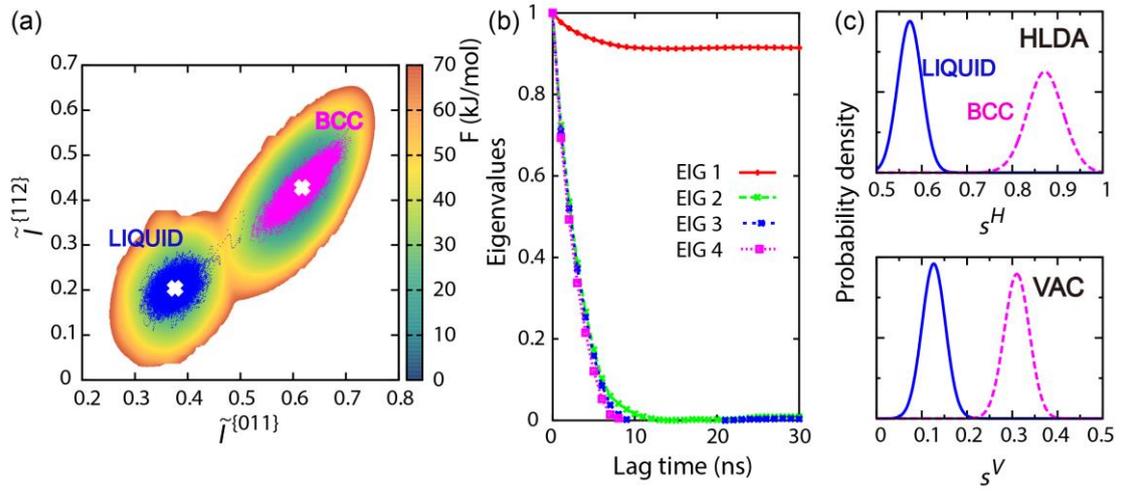

**Figure 2.** (a) FES as a function of $\tilde{I}^{\{011\}}$ and $\tilde{I}^{\{112\}}$ descriptors for the liquid/solid phase transition in Na. The blue and magenta dots represent the points sampled in the unbiased trajectories. The expectation value of each state is marked with a white cross. The projection to other descriptors could be found in Figure S1 in the SI. (b) Evolution of the eigenvalues in VAC as a function of different lag time τ. (c) Probability densities of the unbiased distributions projected onto the HLDA and VAC CVs, with the blue lines and magenta dashed lines corresponding to liquid and BCC phases, respectively.

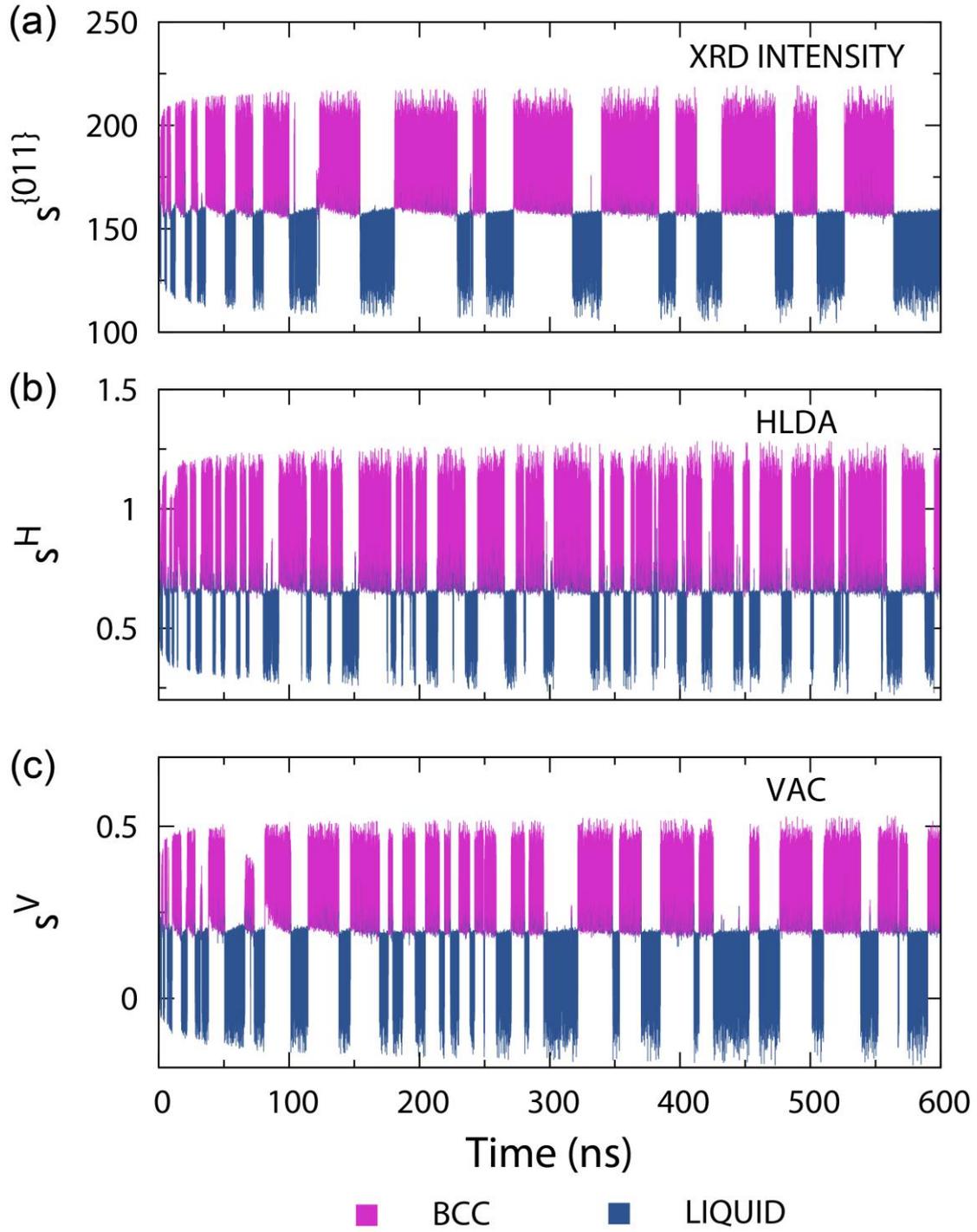

**Figure 3.** Time evolution of the (a) XRD intensity CV $s^{\{011\}}$, (b) HLDA CV $s^H$, and (c) VAC CV $s^V$ in metadynamics simulations for Na.

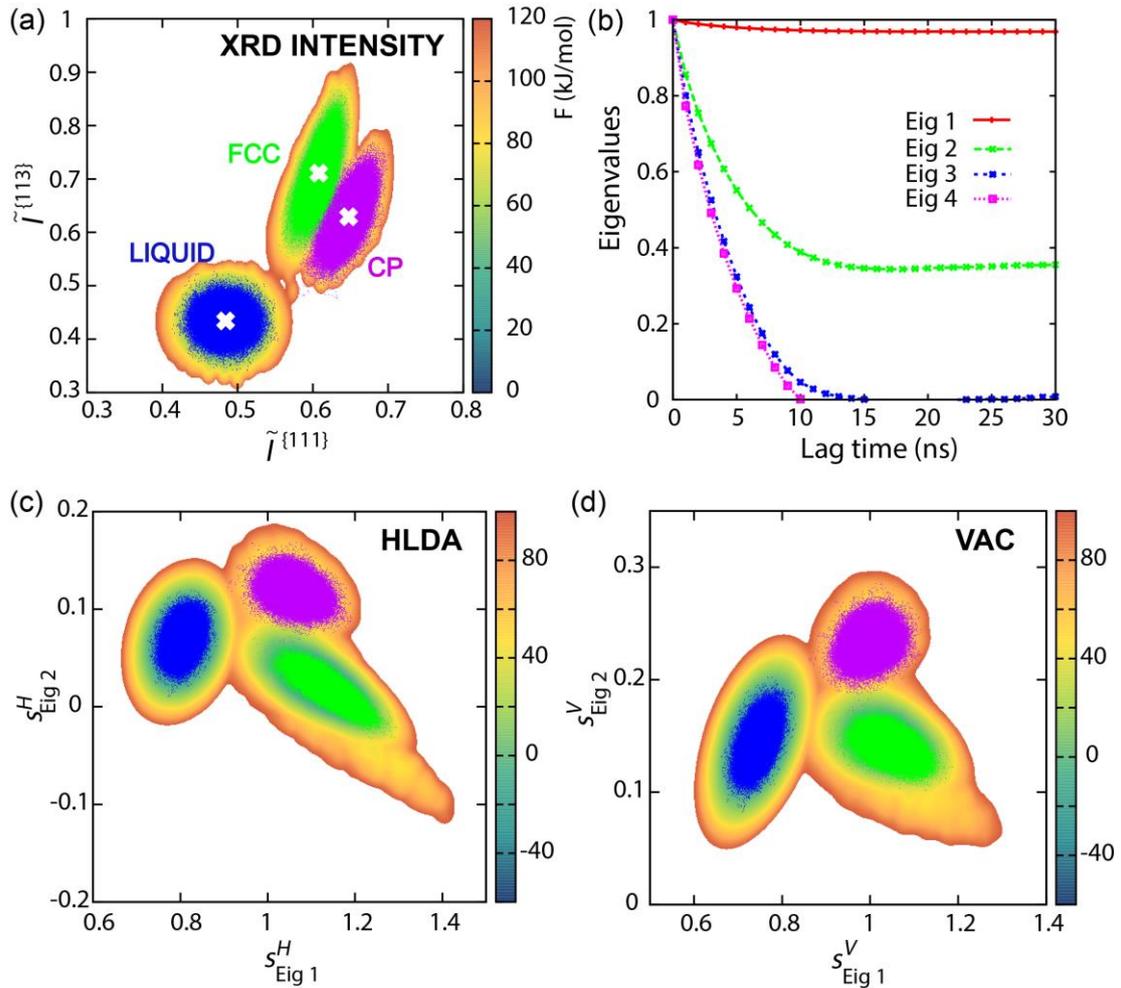

**Figure 4.** (a) FES with respect to descriptors $\tilde{I}^{\{111\}}$ and $\tilde{I}^{\{113\}}$ and points sampled in unbiased runs of Al. Points in liquid, FCC and CP basins are represented by blue, green and magenta dots. The expectation value of each state is marked with a white cross. The projection to other descriptors could be found in Figure S4 in the SI. (b) Evolution of the eigenvalues in VAC as a function of lag time τ. (c,d) FES with respect to two-dimensional HLDA and VAC CVs with points sampled in unbiased runs. The liquid, FCC and CP basins are well separated in both CV spaces.

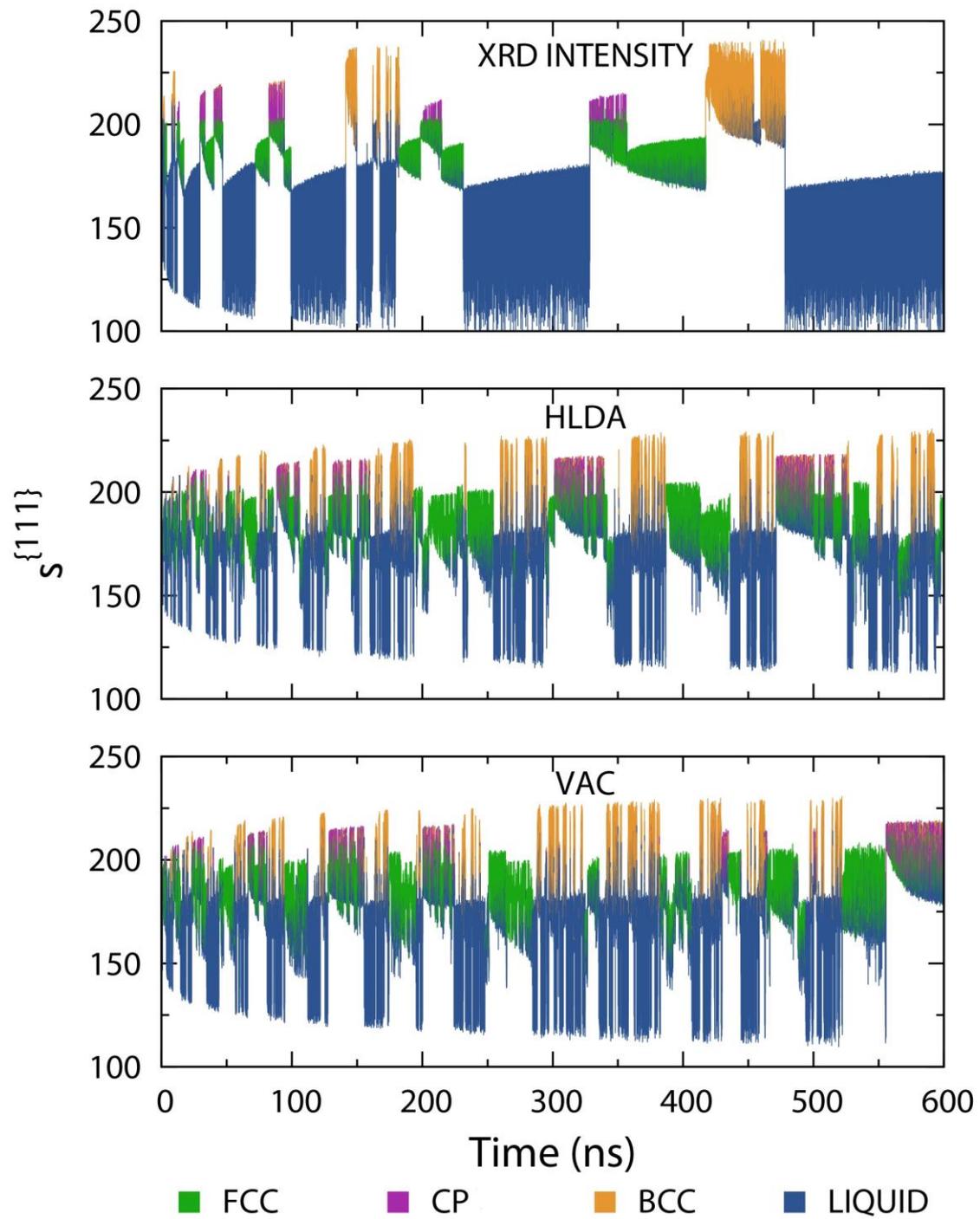

**Figure 5.** Time evolution of XRD intensity $s^{\{111\}}$ in metadynamics simulations based on (a) $s^{\{111\}}$, (b) HLDA CV $s^H$, and (c) VAC CV $s^V$ in Al.

**Table 1.** HLDA and VAC coefficients for the linear combination of XRD intensity peaks for crystallization of Na

|      | $I^{\{011\}}$ | $I^{\{002\}}$ | $I^{\{112\}}$ | $I^{\{022\}}$ |
|------|---------------|---------------|---------------|---------------|
| HLDA | 0.857         | 0.201         | 0.464         | 0.096         |
| VAC  | 0.923         | 0.011         | -0.331        | -0.198        |

**Table 2.** HLDA and VAC coefficients for the linear combination of XRD intensity peaks for crystallization of Al

|      |       | $I^{\{111\}}$ | $I^{\{002\}}$ | $I^{\{022\}}$ | $I^{\{113\}}$ |
|------|-------|---------------|---------------|---------------|---------------|
| HLDA | Eig 1 | 0.946         | 0.028         | 0.077         | 0.315         |
|      | Eig 2 | 0.871         | -0.044        | -0.237        | -0.429        |
| VAC  | Eig 1 | 0.695         | -0.023        | 0.217         | 0.685         |
|      | Eig 2 | 0.932         | -0.087        | -0.220        | -0.273        |